AISI | AI SAFETY INSTITUTE

# Safety Cases: A Scalable Approach to Frontier AI Safety

Benjamin Hilton, Marie Davidsen Buhl, Tomek Korbak, Geoffrey Irving

*Safety cases – clear, assessable arguments for the safety of a system in a given context – are a widely-used technique across various industries for showing a decision-maker (e.g. boards, customers, third parties) that a system is safe. In this paper, we cover how and why frontier AI developers might also want to use safety cases. We then argue that writing and reviewing safety cases would substantially assist in the fulfilment of many of the Frontier AI Safety Commitments. Finally, we outline open research questions on the methodology, implementation, and technical details of safety cases.*

## 1 Introduction

Under the Seoul Frontier AI Safety Commitments (UK Department for Science, Innovation and Technology [DSIT], 2024), sixteen companies agreed to assess risk, identify mitigations, and set out processes to ensure the safety of AI systems they develop. We can split the actions required by these commitments into two broad categories: organisational actions that apply across many AI systems (e.g. setting predefined thresholds), and system-specific actions (e.g. carrying out capability evaluations on a new system).

Many of the system-specific commitments can be fulfilled using safety cases. Safety cases are a way of arguing that a particular system is safe in a given context. Safety cases are beginning to be referred to in signatories' safety frameworks as a way of assuring the safety of new AI systems (Google DeepMind, 2024; Anthropic, 2024a).

Safety cases are used across the automotive (including autonomous vehicles), civil aviation, defence, nuclear, petrochemical, rail and healthcare industries (Favaro et al., 2023; Sujan et al., 2016; Bloomfield et al., 2012; Inge, 2007). While much of the best practice from these industries is likely to be useful in the case of frontier AI, much will have to be tailored to AI's unique circumstances. As a result, we define safety cases by their goal – to be clear and assessable – rather than by specific practices.

1   **Definition 1. Safety case.** A structured argument, supported by a body of evidence, that provides a compelling, comprehensible, and valid case that a system is safe for a given application in a given environment (UK Ministry of Defence [MoD], 2007).

Research is already underway to determine which practices are most useful in the case of AI (Bloomfield et al., 2021; Favaro et al., 2023; Khlaaf, 2023; Clymer et al., 2024; Irving, 2024; Buhl et al., 2024) – including which definition of "safe" is best for various purposes. As we will argue in Section 2, we have a good idea of how to write safety cases for current systems, but we do not yet know how to write good safety cases for future, higher-capability systems. Sketching safety cases can help make gaps in our knowledge apparent.

We will begin by considering when and why safety cases could be used for Frontier AI (Section 2), and then look at their relationship to the Frontier AI Safety Commitments (Section 3). Finally, we will list some open problems on the use of safety cases (Section 4).

## 2 Safety cases for frontier AI systems

This section provides background by outlining how safety cases can be used to inform decision-making (Section 2.1), why doing so might be valuable (Section 2.2), what current safety cases could look like (Section 2.3) and how they might develop in the future (Section 2.4).

### 2.1 A model for the use of safety cases

We can model the various uses of clear and assessable arguments for the safety of a system by considering three parties who could use a safety case (fig. 1):

- A writer develops an argument that a system is safe.
- A red team assesses that argument.
- A decision-maker uses the safety case to decide about the use of the system.

**Figure 1: Diagram showing the use of a safety case**

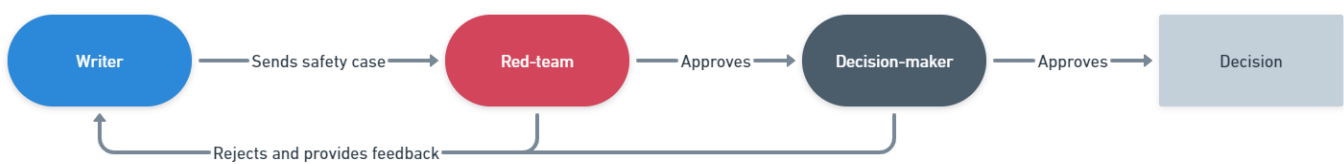

Adapted from Buhl et al., 2024.

Safety cases can be made and used across the lifecycle of an AI system. In other industries, they are usually intended as a 'living document' (Kelly, 1998), kept continuously up to date throughout the operational life of the system. In addition to supporting major "one-off" decisions, such as whether to deploy a system, safety cases can be a useful reasoning tool and source of information that helps developers inform many smaller decisions. For example, safety cases can be used during development to help identify necessary safeguards or after deployment to help assess the impact of new information, such as a new jailbreak.

Possible use cases, and actors who might fill the three roles in each case, are shown in Table 1.

**Table 1: Some examples use cases for safety cases**

| Use | Writer | Red team | Decision-maker |
|---|---|---|---|
| **Improving the design of an AI system, to make it safer** | Developers designing and building the system | Developers designing and building the system, or an independent red team (e.g. in a third-party organisation) | Developers designing and building the system |
| **Internal "go/no-go" decision-making about an AI system (e.g. deciding about training, deployment or post-deployment measures)** | Safety case writing team in an AI company | Safety case red team in an AI company or third-party organisation | Decision-maker in an AI company (e.g. company board) |
| **Demonstrating regulatory compliance for an AI system** | Safety case writing team in an AI company | Regulator or third party | Regulator |
| **Meeting system-specific Frontier AI safety commitments (see table 2)** | Safety case writing team in an AI company | Safety review team in an AI company | Decision-maker in an AI company (e.g. company board) |
| **Responding to incidents, feedback and near misses** | Incident response team | Safety case red team in an AI company or third-party organisation | Decision-maker in an AI company (e.g. company board) |

Safety cases are not always sufficient for actors to fulfil their roles in each of these use cases. Decision-makers may require documents other than just a safety case, for example if they are making decisions based on factors other than safety. The red team would also ideally be empowered to do more than just read the safety case: they should be able to run experiments and produce counterexamples that demonstrate if the claims made by the safety case writer are false. Nevertheless, the key thing that needed to need to carry out all three roles is an argument for the safety of the system that is both correct and assessable – this is a safety case.

Conventionally, safety cases include:

- A top-level claim, which specifies what is meant by "safe", and details the conditions under which the safety case holds (for example use cases or use environments).

- A structured argument, often written as a graph, linking this top-level claim to subclaims and, eventually, to evidence. This structured argument provides a principled backbone of an overall narrative about why the system is safe. The argument is often provided in a "safety case report" (the highest-level document in a full safety case) that also discusses the assumptions and details behind the argument.

- The evidence that supports the argument (for example, results from testing a system).

However, we remain uncertain what practices from other industries would be best for the frontier-AI use cases described in this section.

## 2.2 Why use safety cases?

We can contrast safety cases against several other approaches for ensuring that systems are safe:

- Prescriptive approaches that specify precise safety requirements in advance of developing a system (e.g. publicly pre-committing to specific safety mitigations, for all new AI systems, in response to specific capability thresholds that those systems might cross).

- Lighter-touch approaches that primarily rely on information-sharing (e.g. sharing various facts and evidence about the system, without structuring it as an argument).

- *Ex post* approaches that investigate and correct safety failures as they materialise (e.g. accident investigation, safety audits, continuous monitoring; these could also form parts of other approaches – the distinguishing feature of *ex post* approaches is the lack of an *ex ante* assessment of the safety of a system).

Comparing safety cases to prescriptive approaches, the key distinguishing feature of safety case is flexibility: the safety case writer can deviate substantially from standard practices as long the reviewers end up confident that the resulting argument is correct. For example, the safety case writer may include a completely novel argument for system safety in their safety case. Safety cases' flexibility brings several advantages over prescriptive approaches (Buhl et al., 2024):

- It can be hard to specify requirements in advance. Frontier AI systems have a wide range of use cases and use environments, and broad disagreement about the extent of the risks, which makes it difficult to pre-specify which actions will be needed for a given system. More importantly, frontier AI systems are changing quickly – of particular relevance is the rapidly developing field of designing and evaluating safety mitigations (Casper et al., 2023; Greenblatt et al., 2024). Safety cases offer a more durable approach, at least until more specific standards are developed and agreed.

- If discovered either that any prescriptive requirements are insufficient to ensure the safety of a system, or that a system does not meet the requirements but can

- nevertheless be considered safe for other reasons, then that system can still pass an assessment using a safety-case approach.

- Assuming they are written and assessed correctly, safety cases are also more robust: a safety case could consist of a set of requirements, but it would also need to argue that these requirements are sufficient for safety.

Comparing safety cases to lighter-touch approaches, the key distinguishing feature of safety cases is their explicit argumentation – which allows for increased clarity and assessability. Explicit argumentation can help both the writing team and the red team identify flaws in existing safety practices by forcing them to consider what assumptions are being made in otherwise implicit arguments. It can also make it easier to identify sources of disagreement. Finally, writing safety cases for present systems can help identify assumptions that may not hold for future systems, such that further research is needed. Sketching safety case arguments for future systems can have a similar function.

Comparing safety cases to *ex post* approaches, the key distinguishing feature of safety cases is that they require upfront consideration of safety. *Ex post* approaches have lower upfront costs and can be suitable in circumstances where risks are difficult to identify in advance. However, they are generally more prone to mistakes – as, by definition, the decision to deploy a system cannot be checked *ex ante*. Even if risks are mitigated once they are detected, there will be periods of time when there is high *ex ante* risk, even if that risk should not have been taken. As a result, the cost savings from *ex post* approaches are more justifiable when the expected cost of this additional assumed risk is lower than these savings. These savings also seem more likely to be justifiable when risks lead to reversible damages (e.g. economic costs that can be repaid) than irreversible damages (e.g. loss of life).

There are downsides to safety cases. Most importantly, the robustness of safety cases, as compared to lighter-touch approaches, implies a necessarily more intensive process – as we will see below, for current AI systems, this extra cost could be small, but for future systems we may need much more complex or expensive evidence and argumentation. These complex safety cases will then only be a good solution if we are building high-risk and capital-intensive systems; this is the case for many of the industries where safety cases are currently used. We are uncertain about the capital costs and risks posed by future frontier AI systems. However, some trends suggest continued investment and capability increase (Cottier et al., 2024).

## 2.3 The possible content of safety cases for current systems

For many systems, we think that they are safe because they are not capable of doing anything critically dangerous – and the appropriate capability evaluations can show this. This is the implicit argument behind many published safety frameworks (OpenAI, 2023; Google DeepMind, 2024; Anthropic, 2024b) and has been called an "inability" argument (Clymer et al., 2024).

A safety case for a low-capability system would make this argument explicit, bringing together capability evaluation results alongside a risk assessment. Writing this argument as part of a safety case would help ensure that the evaluations carried out do indeed show that the system is safe and help identify areas where more might need to be added to the safety case in the future if capability thresholds are passed.

A safety case based on the argument sketched above would need to justify two key claims. First, that an appropriate set of risk models and capability thresholds have been considered, such that if they are not a concern, the system is safe enough overall. Second, that the capability evaluations have been carried out to a high enough standard that we can be confident the system does not possess the relevant capabilities. Goemans et al. (2024) provides an example safety case template for an inability argument.

**Figure 2: A possible structure for an inability argument**

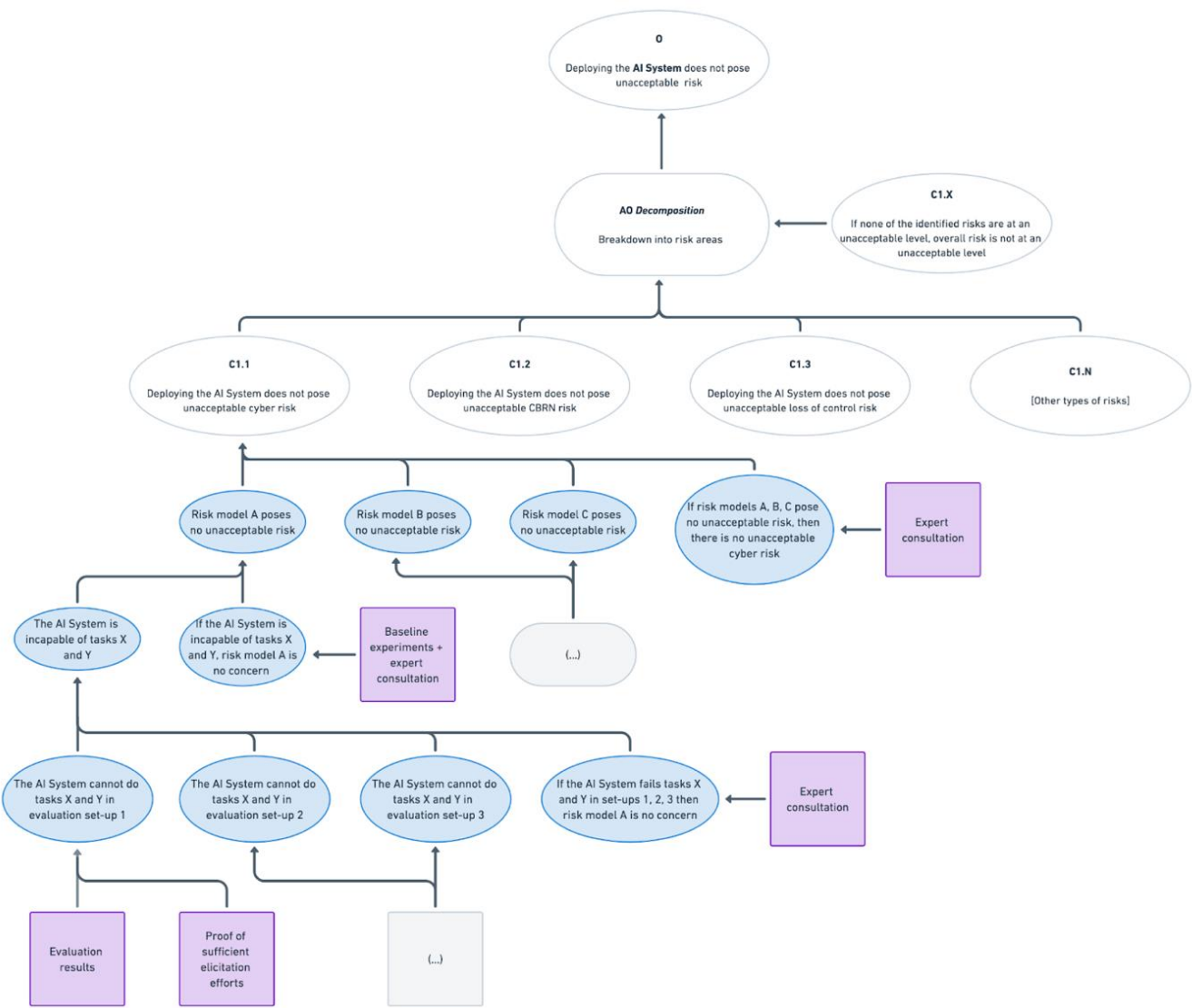

Adapted from Goemans et al., 2024. The argument for claim C1.1 is presented here in summarised form; see Goemans et al., 2024 for a detailed inability safety case template.

A safety case for a low-capability system would ideally also include governance arguments: for example, it might argue that there is an appropriate organisational safety culture, or that relevant staff have appropriate knowledge and training.

As the results of capability evaluations approach thresholds beyond which we think that systems should be considered unsafe, our confidence in an inability argument should decrease. We should be uncertain about the extent to which the evaluations represent true risky capabilities, about the validity of our thresholds (in particular, whether the system could be unsafe even if below these thresholds), and about other facts about the system, such as the possibility of sandbagging ([van der Weij, 2024](#)). These uncertainties grow as capabilities advance.

As a result, for many future systems, safety cases will require further kinds of arguments. One kind focuses on misuse safeguards: arguing that while a model is capable of causing risk in some misuse scenarios, there are sufficient safeguards around the model that prevent the system being misused in this way (a "safeguards" argument).

## 2.4 What might safety cases look like for future systems?

For future, increasingly capable systems, we may no longer be confident in any ways of making inability arguments. Similarly, it is likely that, if systems become more capable, it will become more difficult to build sufficient safeguards – so arguments beyond inability and safeguards will be needed. By layering many independent arguments, we can achieve a higher degree of confidence in the overall claim that the system is safe.

Further kinds of arguments may respond to the specific problems posed by *autonomous* systems. An "AI control" argument ([Greenblatt et al., 2023](#)) makes the case that an AI system has sufficient measures to prevent autonomous harmful activity, even when the model is actively trying to subvert those measures. A "trustworthiness" argument makes the case that even if an AI system is capable of causing harm, and the safeguards (against misuse) and control measures (against autonomous risks) are insufficient to prevent this, the system nevertheless will not cause harm – for example, because the system is aligned ([Clymer at al., 2024](#)).

We do not currently know how to make these arguments well. Figuring this out will require technical machine learning research alongside learning from other industries. But this is not a problem unique to a safety cases approach. Our lack of knowledge is the same under any approach; safety cases just make the gaps apparent. And we can start this research now – some organisations, including UK AISI, have begun doing so ([Bloomfield et al., 2021](#); [UL, 2023](#); [Correa-Jullian et al., 2023](#); [Khlaaf, 2023](#); [Clymer et al., 2024](#); [Irving, 2024](#)).

# 3 How safety cases can help fulfil the Frontier AI Safety Commitments

Signatories of the Frontier AI Safety Commitments ([UK Department for Science, Innovation and Technology [DSIT], 2024](#)) commit to develop and deploy their frontier AI models and systems responsibly, and to demonstrate how they have achieved this by publishing a safety framework. This section discusses how safety cases relate to safety frameworks (Section 3.1), before examining each commitment in turn, looking at the extent to which safety cases can be used to meet that commitment (Section 3.2).

## 3.1 Relationship to safety frameworks

Safety frameworks and safety cases are complementary and serve different roles, summarised in Table 2.

**Table 2: The role of safety frameworks and safety cases**

| Role of safety *frameworks* | Role of safety *cases* |
|---|---|
| • Ensure adherence to the commitments at the *organisational* level, including wider policies and culture<br>• Amortise costs by conducting analysis that applies to many systems<br>• Prepare for risks ahead of time<br>• Facilitate public discussion, critique and improvements of safety framework<br>• Set out roles and responsibilities | • Ensure adherence to the commitments at the *system* level<br>• Adapt analysis to the details of a specific system<br>• Hold relevant actors (e.g. developers, monitors, red teams) accountable for adequate implementation<br>• Red-team safety framework implementation |

Safety frameworks and safety cases are interrelated. In particular:

- Safety cases could use adherence to a safety framework as part of their central argument. For example, a safety case could argue that the model is below the capability thresholds set out in the safety framework, or that the implemented mitigations are adequate. Since capability thresholds and safety mitigations are rarely fully specified in safety frameworks, substantial argumentation may be required – for which safety cases are a useful tool. However, safety cases maintain the flexibility to use other arguments if the safety framework is not suitable (e.g. if new information has appeared since the writing of the safety framework, or if there are unique circumstances with a particular system to which the framework does not apply).

- Safety frameworks could include a commitment to produce safety cases. They could also set out a 'safety case process', explaining when safety cases will be produced, as well as how they will be red-teamed, updated, and used in decision-making. This provides broader assurance than pre-committing to a specific set of evaluations and safeguards, in that the developer *also* commits to check a specific system at specific times against those pre-commitments.

The ideal safety case process remains uncertain – this will depend on the details about the contents of the safety cases which, as we note above, is an active area of research. That said, safety case processes in other industries tend to include (Office for Nuclear Regulation [ONR], 2022):

- Details of the steps for writing a safety case.

- Ways writing a safety might go wrong, and ways to prevent that.
- Monitoring and testing of the safety case writing process.
- Feedback mechanisms for reviewing both safety cases themselves and weak points in the safety case writing process.
- Roles and responsibilities for those involved in the process as well as definitions of the training and qualifications required for these roles.
- Additional measures during times of high stress (e.g. tight deadlines, intense commercial or operational pressure).

Developers can already start setting out a careful and comprehensive safety case process in their safety frameworks, sharing safety cases with third parties, and publishing adjacent documents to help build trust with downstream developers, research partners, governments and the general public.

## 3.2 Relationship to individual commitments

The Frontier AI Safety Commitments include eight sub-commitments, organised around three outcomes: (1) effective risk identification, assessment, and mitigation, (2) accountability and governance, and (3) transparency (UK Department for Science, Innovation and Technology [DSIT], 2024). Table 3 outlines how safety cases relate to each individual sub-commitment. We assume that safety cases are used in combination with a safety framework which specifies a safety case process, as described in the previous section.

**Table 3: The extent to which safety cases can, by themselves, meet the Frontier AI Safety Commitments**

| Commitment | *If a developer specifies and executes an adequate safety case process that produces correct safety cases…* | |
| --- | --- | --- |
| | **How does this help meet this commitment?** | **What more, beyond safety cases, is needed to meet this commitment?** |
| **I (Risk assessment)**<br><br>*"Assess the risks posed by their frontier models or systems across the AI lifecycle, including before deploying that model or system, and, as appropriate, before and during training. Risk assessments should consider model capabilities and the context in which they are developed and deployed, as well as the efficacy of implemented mitigations to reduce the risks associated with their foreseeable use and misuse. They should also consider results from internal and external evaluations as appropriate, such as by independent third-party evaluators, their home governments, and other bodies their governments deem appropriate."* | A hazard analysis will usually form a crucial part of a safety case. For example, a hazard identification procedure is needed to identify necessary evaluations for an inability argument. The results of these evaluations then feed into a specific risk assessment. | An adequate safety case process could in theory be sufficient.<br><br>However, safety frameworks should still include a hazard analysis to ensure preparedness and facilitate scrutiny.<br><br>Additionally, the cost of the hazard analysis could be amortised as it may not change substantially for new systems. |
| **II (Identify and assess thresholds)**<br><br>*"Set out thresholds at which severe risks posed by a model or system, unless adequately mitigated, would be deemed intolerable. Assess whether these thresholds have been breached, including monitoring how close a model or system is to such a breach. These thresholds should be defined with input from trusted actors, including organisations' respective home governments as appropriate. They should align with relevant international agreements to which their home governments are party. They should also be accompanied by an explanation of how thresholds were decided upon, and by specific examples of situations where the models or systems would pose intolerable risk."* | A safety case should assess whether thresholds have been breached. | Safety cases make use of thresholds, but a separate process is needed to define the thresholds and why they are appropriate.<br><br>Thresholds can be breached at any time, including during training and after deployment. While safety cases can be produced and updated throughout the lifecycle, more easily measurable triggers and rapid response plans will likely be needed given the risk of sudden capability jumps or incidents. These processes could be specified in safety frameworks and/or pre-training safety cases. |
| **III (Mitigations)**<br><br>*"Articulate how risk mitigations will be identified and implemented to keep risks within defined thresholds, including safety and security-related risk mitigations such as modifying system behaviours and implementing robust security controls for unreleased model weights."* | The adequacy of mitigations should be argued for in a safety case; identifying mitigations will be necessary to do this. | An adequate safety case process could in theory be sufficient.<br><br>However, safety frameworks should likely still include best guesses about necessary mitigations, to ensure preparedness and facilitate scrutiny. |
| **IV (Actions when thresholds are exceeded)**<br><br>*"Set out explicit processes they intend to follow if their model or system poses risks that meet or exceed the pre-defined thresholds. This includes processes to further develop and deploy their systems and models only if they assess that residual risks would stay below the thresholds. In the extreme, organisations commit not to develop or deploy a model or system at all, if mitigations cannot be applied to keep risks below the thresholds."* | A safety case should include a description of what happens if the safety case is invalidated (for example if thresholds are breached after deployment). | Organisational actions, such as a commitment not to develop or deploy certain systems, would not be included in safety cases. However, this commitment could be phrased (e.g. in a safety framework) as a commitment not to develop or deploy systems without writing and assessing a correct safety case. |
| **V (Continual improvements)**<br><br>*"Continually invest in advancing their ability to implement commitments i-iv, including risk assessment and identification, thresholds definition, and mitigation effectiveness. This should include processes to assess and monitor the adequacy of mitigations, and identify additional mitigations as needed to ensure risks remain below the pre-defined thresholds. They will contribute to and take into account emerging best practice, international standards, and science on AI risk identification, assessment, and mitigation."* | Safety cases could help identify areas of research needed to ensure safety for future systems. | Meeting this commitment would also require investment in producing safety cases effectively, monitoring the techniques used, and ensuring that the process used to write and assess safety cases is kept up to date. |
| **VI (Accountability)**<br><br>*"Adhere to the commitments outlined in I-V, including by developing and continuously reviewing internal accountability and governance frameworks and assigning roles, responsibilities and sufficient resources to do so."* | A safety case process could constitute an important accountability mechanism, especially if it includes rigorous red teaming and governance mechanisms. | Broader processes (e.g. governance audits, rehearsals of emergency response plans) will likely be needed in addition to the system-specific reviews provided by safety cases. |
| **VII (Transparency)**<br><br>*"Provide public transparency on the implementation of the above (I-VI), except insofar as doing so would increase risk or divulge sensitive commercial information to a degree disproportionate to the societal benefit. They should still share more detailed information which cannot be shared publicly with trusted actors, including their respective home governments or appointed body, as appropriate."* | A safety case could be shared with trusted government bodies and third parties. | It would be difficult to publicly publish full safety cases as they will usually contain sensitive information. Other adjacent documents, such as safety frameworks, will need to be prepared to share with the public. |
| **VII (Involvement of external actors)**<br><br>*"Explain how, if at all, external actors, such as governments, civil society, academics, and the public are involved in the process of assessing the risks of their AI models and systems, the adequacy of their safety framework (as described under I-VI), and their adherence to that framework."* | External actors could be involved with assessing or even writing safety cases (although the ecosystem to support this is nascent). | External actors may also need to be involved with the development of safety frameworks and broader governance mechanisms (e.g. governance audits). |

## 4 Open problems about safety cases for frontier AI

Throughout this paper, we have noted our uncertainty about what safety cases and safety case processes for frontier AI systems should look like. This section contains a highly non-comprehensive list of research questions, in the hope that answering these could improve our ability to write safety cases for frontier AI. We begin with foundational questions about how to write good safety cases (Section 4.1), then discuss implementation challenges (Section 4.2), and finally outline the technical AI safety research that might be needed to be able to write safety cases for high-capability systems (Section 4.3).

The UK AI Safety Institute is aiming to act as a hub for research on safety case sketches (Irving, 2024) and safety case templates (UK AI Safety Institute [AISI], 2024): safety case sketches detail the arguments and evidence we expect for specific safety methods for a particular system; safety case templates are rough arguments with details that can be filled in for a variety of possible system. It can be hard to answer questions about the best ways of writing safety cases in the abstract. Getting into the details through sketches and templates tests our hypotheses about the best ways of writing safety cases and completed sketches and templates could also form indicative building blocks for full safety cases in the future.

### 4.1 Writing good safety cases

While there is much to be learned from other industries, we will need to adapt best practice to the unique situation of frontier AI. Some of the initial questions in this space include:

- **Top-level claims.** It is not clear how the top-level claim in a safety case should be specified – in part because there is no broad agreement on how low risk needs to be for sufficient safety, and in part because of difficulty even phrasing these claims. While principles like 'ALARP' (as low as reasonably practicable) (Melchers, 2001), 'GALE' (globally at least equivalent to) (Salako et al., 2021) and 'ACARP' (as confident as reasonably practicable) (Bloomfield and Bishop, 2009) have been used historically, frontier AI may require new principles. What sort of top-level claims would be ideal for frontier AI?

- **Argument types.** Safety cases are stronger if they contain multiple independent parallel arguments. What broad argument types should be included in frontier AI safety cases? How can we tell the extent to which these arguments are independent?

    o **Sociotechnical arguments.** Frontier AI safety cases should make arguments about more than just the technical system. How should sociotechnical arguments – for example arguments about organisational governance – feature in a safety case?

- **Notation.** A variety of different forms of notation have been used to show the structure of the argument in a safety case; common notation schemes include Claims, Arguments, Evidence (CAE) and Goal Structuring Notation (GSN) (Kelly, 2007). Which notation scheme works best for frontier AI?

- **Quantification.** In software safety cases, it can be hard to find objective quantitative measures for the confidence (Bloomfield and Rushby, 2021). This seems particularly hard for AI systems – for example, because it is hard to measure the rate of failure of components. Should we therefore propagate *subjective* probabilities throughout a safety case? Are there alternative approaches that could be more assessable by a red-team or decision-maker?

- **Generalisability.** It is often not possible to create arguments and evaluations that explicitly cover every potential deployment scenario and use case. How much do safety case arguments and evidence generalise to other deployment scenarios and use cases?

## 4.2 Implementation challenges

Beyond questions about how to write good safety cases, there are important questions about how safety cases should be used in practice. Some key questions include:

- **Safety case processes.** Safety cases need both effective writing and assessment processes. This includes defining when safety cases should be written and updated, as well as determining the appropriate level of independence and frequency for red-team assessment. What is the ideal end-to-end process for developing and red-teaming safety cases?

- **Organisational integration.** Safety case processes must fit within existing organisational structures and governance frameworks. This includes establishing clear lines of accountability and ensuring appropriate expertise. What qualifications or training should be required for safety case writers and reviewers? How should they work with relevant other actors such as those running evaluations, safeguarding against jailbreaks, or writing safety frameworks?

- **Ensuring continued validity of a safety case.** Safety cases can fail in several ways: something about the operational environment might change, making the safety case no longer correct, or we could gain evidence that the safety case was initially wrong (e.g. a risk materialises, or an assessment of the safety case finds a flaw). Organisations will need clear protocols for these scenarios. How should developers and others monitor if a safety case continues to be valid and what should happen when a safety case fails?

These questions will become more important as organisations begin implementing safety case processes. While we can draw on experience from other industries, practical experimentation with different approaches will be helpful for developing best practices.

## 4.3 Gaps in technical knowledge

Many safety case sketches cannot be fleshed out into full safety cases today because providing evidence supporting their assumptions is blocked by an open research problem. Some of these open problems include:

- **Capability elicitation.** Safety cases often rely on evidence that a model cannot do certain dangerous things. The best evidence for this is usually that competent red teams fail to elicit a capability despite their best efforts. However, this falls short of showing that no one could elicit the capability. While supervised finetuning on optimal demonstrations may help prevent under-elicitation (Greenblatt et al., 2024), more research is needed to conclude that supervised finetuning can elicit most of an LLM's capabilities. How robust is supervised finetuning as a capability elicitation technique, and what other techniques could strengthen a safety case?

- **System reasoning.** Many possible safety cases could rely on an improved understanding of how models reason. For instance, how good are LLMs at reasoning without chain-of-thought prompting (Wang et al., 2024)?

- **Scheming.** It will be harder to build safety cases for models exhibiting scheming behaviour (understood as strategically deceiving overseers in pursuing goals; Carlsmith, 2023). Such behaviour could necessitate stringent AI control measures (Greenblatt et al., 2024). Progress in understanding this behaviour could come from studying model organisms (Hubinger et al., 2023) – for example, deliberately prompting LLMs to scheme and analysing how their propensity to scheme depends on various factors (Scheurer et al., 2023). To what extent do current training pipelines incentivize deceptive behaviour (Denison et al., 2024)? And how can we measure capabilities and ensure safeguard effectiveness for scheming systems?

- **Generalisation.** Many safety cases depend on monitoring systems continuing to work as models scale or their input distribution changes. This includes both models monitoring outputs, linear probes, or mechanistic interpretability tools like sparse autoencoders. We need better ways to assess how well these generalise and how resistant they are to circumvention (Sharkey, 2022). How can we better estimate the out-of-distribution generalisation of monitors?

Progress on these technical problems will be crucial for writing convincing safety cases for more capable AI systems. However, work on safety case methodology need not wait for these problems to be solved - developing better safety case processes now will help identify which technical problems are most important to solve.

# 5 Conclusions

Safety cases are a useful and important tool for reasoning and communicating about system safety in a way that is clear and assessable.

They have several advantages – in addition to being clear and assessable, they are flexible, focused, and helpful for identifying where new arguments will be required for future systems. This may make them more robust and scalable than requirements-based approaches.

Safety cases for low capability systems may primarily require bringing together work that AI developers are already doing (a risk assessment and capability evaluations) into a structured document. As a result, they could begin being used today as a way of fulfilling many of the Frontier AI Safety Commitments. This would help build the infrastructure and practice of writing safety cases, as well as helping to identify the open research problems we'll need to close before we can successfully argue for the safety of future systems.

19